\documentclass[aoas,nameyear,seceqn,rotating,dvips]{arximspdf}
\usepackage{dcolumn}
\usepackage{graphicx}

\doi{10.1214/09-AOAS296}
\volume{4}
\issue{2}
\pubyear{2010}
\firstpage{916}
\lastpage{942}

\makeatletter

\newcolumntype{d}[1]{D{.}{.}{#1}}

\def\eqref#1{(\ref{#1})}

\makeatother

\begin{document}
\begin{frontmatter}

\title{A Dirichlet process mixture of hidden Markov models for protein
structure prediction\protect\thanksref{T1}}
\runtitle{A DPM--HMM for protein structure prediction}
\thankstext{T1}{Supported in part by NIH/NIGMS Grant R01GM81631.}

\begin{aug}
\author[a]{\fnms{Kristin P.} \snm{Lennox}\ead[label=e1]{lennox@stat.tamu.edu}\corref{}},
\author[a]{\fnms{David B.} \snm{Dahl}\ead[label=e2]{dahl@stat.tamu.edu}},
\author[b]{\fnms{Marina} \snm{Vannucci}\ead[label=e3]{marina@rice.edu}\thanksref{T2}},
\author[c]{\fnms{Ryan}~\snm{Day}\ead[label=e4]{rday@pacific.edu}}
\and
\author[c]{\fnms{Jerry W.} \snm{Tsai}\ead[label=e5]{jtsai@pacific.edu}}
\thankstext{T2}{Supported in part by NIH/NHGRI Grant R01HG003319 and by NSF/DMS
Grant DMS-06-05001.}
\runauthor{K. Lennox et al.}
\affiliation{Texas A\&M University, Texas A\&M University, Rice
University,\\ University of the Pacific and University of the Pacific}
\address[a]{K. P. Lennox\\ D. B. Dahl\\ Department of Statistics\\
Texas A\&M University\\
3143 TAMU\\ College Station, Texas 77843-3143\\
USA\\ \printead{e1}\\
\phantom{E-mail: }\printead*{e2}}

\address[b]{M. Vannucci\\ Department of Statistics\\
Rice University\\
MS 138\\ Houston, Texas 77251-1892\\ USA\\
\printead{e3}}

\address[c]{R. Day\\ J. W. Tsai\\ Department of Chemistry\\
University of the Pacific\\
3601 Pacific Ave\\
Stockton, California 95211-0110\\
USA\\ \printead{e4}\\
\phantom{E-mail: }\printead*{e5}}
\end{aug}

\received{\smonth{9} \syear{2009}}

%
\begin{abstract}
By providing new insights into the distribution of a protein's torsion angles,
recent statistical models for this data have pointed the way to more
efficient methods
for protein structure prediction. Most current approaches have
concentrated on bivariate
models at a single sequence position. There is, however, considerable
value in
simultaneously modeling angle pairs at multiple sequence positions in
a protein.
One area of application for such models is in structure prediction for
the highly
variable loop and turn regions. Such modeling is difficult due to the
fact that
the number of known protein structures available to estimate these
torsion angle
distributions is typically small. Furthermore, the data is ``sparse''
in that not
all proteins have angle pairs at each sequence position. We propose a new
semiparametric model for the joint distributions of angle pairs at multiple
sequence positions. Our model accommodates sparse data by leveraging known
information about the behavior of protein secondary structure. We demonstrate
our technique by predicting the torsion angles in a loop from the
globin fold family.
Our results show that a template-based approach can now be successfully
extended to
modeling the notoriously difficult loop and turn regions.

\end{abstract}

%

\end{frontmatter}

\section{Introduction}
The field of protein structure prediction has greatly benefitted from
formal statistical modeling of available data [\citet
{Osguthorpe2000}; \citet{Bonneau2001}]. More automatic methods for predicting
protein structure are critical in the biological sciences as they help
to overcome a major bottleneck in effectively interpreting and using
the vast amount of genomic information: determining the structure, and
therefore the function, of a gene's protein product. Currently the
growth of genomic data far outstrips the rate at which experimental
methods can solve protein structures. To help accelerate the process,
protein structure prediction methods aim to construct accurate
three-dimensional models of a target protein's native state using only
the protein's amino acid sequence.

Protein structure is typically described in terms of four categories:
primary through quarternary. Primary structure consists of the linear
sequence of covalently bonded amino acids that make up a protein's
polypeptide chain. Secondary structure describes the regularly
repeating local motifs of $\alpha$-helices, $\beta$-strands, turns and
coil regions. For a single polypeptide chain, tertiary structure
describes how the secondary structure elements arrange in
three-dimensional space to define a protein's fold. By allowing the
polypeptide chain to come back on itself, the loops and turns
effectively define the arrangement of the more regular secondary
structure of $\alpha$-helices and $\beta$-strands. Quarternary
structure describes how multiple folded polypeptide chains interact
with one another. In a typical structure prediction problem the primary
structure is known, and the goal is to use this information to predict
the tertiary structure.

One of the standard approaches to this problem is template-based
modeling. Template-based approaches are used when the target sequence
is similar to the sequence of one or more proteins with known
structure, essentially forming a protein fold ``family.'' Typically the
core of the modeled fold is well defined by regular secondary structure
elements. One of the major problems is modeling the loops and turns:
those regions that allow the protein's tertiary structure to circle
back on itself. Unlike the consistency of the core in a template-based
prediction, the variation in the loops and turns (both in terms of
length and amino acid composition) between structures with the same
fold family is often quite large. For this reason current
knowledge-based methods do not use fold family data. Instead of the
template-based approach, they use libraries of loops which are similar
in terms of length and amino acid sequence to the target. However, such
library data sets do not have the same level of structural similarity
as do purely within-family data sets. In this work, our approach to
modeling structural data allows us to effectively extend template-based
modeling to the loop and turn regions and thereby make more informed
predictions of protein structure.

\begin{figure}[t]

\includegraphics{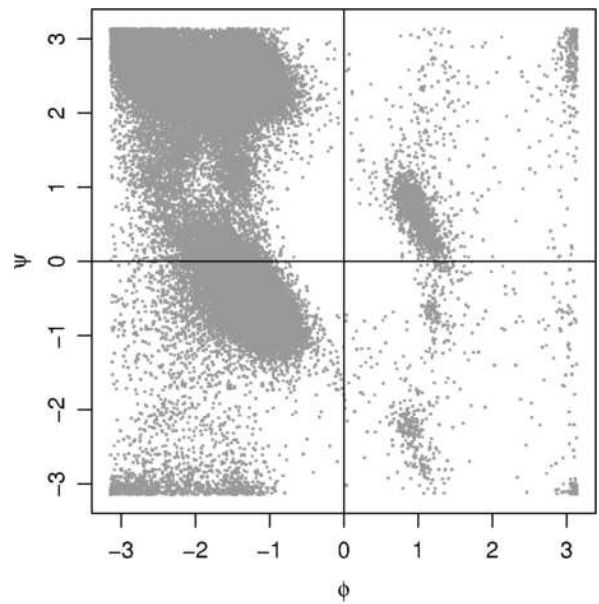}

\caption{Ramachandran plot for the 130,965 angle pairs that make up the
$\mathit{PDB}$ data set for the amino acid alanine. Angles are measured in radians.}
\label{fig:ala}
\end{figure}

Our approach is based on the simplest representation of protein
structure: the so-called backbone torsion angles. This representation
consists of a $(\phi,\psi)$ angle pair at each sequence position in a
protein, and it provides a reduction in complexity from using the 12
Cartesian coordinates for the 4 heavy backbone atoms at each position.
This method for describing protein structure was originally proposed by
\citet{Ramachandran1963}, and the customary graphical representation of
this type of data is the Ramachandran plot. The Ramachandran plot in
Figure \ref{fig:ala} shows the $(\phi,\psi)$ angles of protein
positions containing the amino acid alanine. The pictured data set was
obtained from the Protein Data Bank [PDB, \citet{Kouranov2006}], a
repository of solved protein structures.

Density estimation of Ramachandran space is particularly useful
for\break
template-based structure prediction. Because a target protein with
unknown tertiary structure is known to be related to several proteins
with solved structures, models for bivariate angular data can be used
to estimate the distribution of $(\phi,\psi)$ angles for a protein
family, and thereby generate candidate structures for the target protein.

While there has been considerable recent work on modeling in
Ramachandran space at a single sequence position [see, e.g., \citet
{Ho2003}; \citet{Lovell2003}; \citet{Butterfoss2005};
\citeauthor{Lennox2009b} (\citeyear{Lennox2009b,Lennox2009})], models that
accommodate multiple sequence positions remain uncommon. A notable
exception is the DBN-torus method of \citet{Boomsma2008}. However, this
approach was developed primarily to address sampling of fragments in
{de novo} protein structure prediction, and so specifically does
not include protein family information. {De novo} structure
prediction is used when similar proteins with known structure are
unavailable and is thus inherently more difficult and less accurate
than template
based modeling. While template-based methods can draw on a certain
amount of known information, a common complication is that protein
families typically have fewer than 100 members, and often fewer than 30
members.

Not only do protein families tend to have few members, but the data
within a family is ``sparse,'' particularly in loop regions. A template
sequence for a protein structure family is generated by simultaneously
aligning all of the member proteins using amino acid type at each
sequence position. However, the sequences in a fold family are usually
of different lengths due to different sizes of loops and turns. In such
an alignment, a typical member protein is not represented at every
sequence position. This leads to what we call a ``sparse data''
problem. Note that this is not a missing data situation, as a sequence
position is not merely unobserved, but rather does not in fact exist.

A joint model for a large number of torsion angles using somewhat
limited data can be enhanced by leveraging prior knowledge about the
underlying structure of the data. We present a Bayesian nonparametric
model incorporating a Dirichlet process (DP) with one of two possible
families of centering distributions for modeling the joint
distributions of multiple angle pairs in a protein backbone. Our model
addresses the sparse data situation, and also accommodates a larger
number of sequence positions than previously considered methods of
template-based density estimation. One of our proposed centering
distributions leads to a largely noninformative prior, but we also
propose a family of centering distributions based on known
characteristics of protein secondary structure in the form of a hidden
Markov model (HMM). The inclusion of an HMM allows our model to share
structural information across sequence positions. Since each secondary
structure type has a distinctive footprint on the Ramachandran plot,
with this process we can use an informative prior to incorporate
additional information into our model.

There is precedent for the use of a hidden Markov model for protein
structure prediction in the DBN-torus model of \citet{Boomsma2008}.
There, secondary
structure information is incorporated into the state space of a dynamic Bayesian
network, a generalization of an HMM, which allows the DBN-torus model
to infer secondary
structure when generating candidate angle pair sequences. The model generates
significantly better candidates, however, when secondary structure is
provided from an external secondary structure prediction method. There
are other differences between the DBN-torus method and our own which
result from the distinct applications of the two methods. DBN-torus is
used for {de novo} structure prediction; it is
designed to make predictions for any kind of protein, and is not
customized for
a particular fold family. In contrast, our method is tailored for
template-based modeling. Thus, the DBN-torus model can be used even
when template information is unavailable, but will miss opportunities
for improvement when fold-family structure information exists.

In this paper we apply our method to the loop region between the E and
F $\alpha$-helices of the globin protein template, which varies between
8 and 14 sequence positions in length. By borrowing strength from
neighbors containing numerous observations, our model generates
informative density estimates even if relatively little data is
available at a given position. This property gives our method a
significant advantage in loop prediction by allowing the use of fold
family data. This extension of template-based modeling to loop regions
was not possible before the development of these statistical tools. We
show that using our Dirichlet process mixture of hidden Markov models
(DPM--HMM) in a template-based approach provides a better match to real
structure data than does either a library-based method or DBN-torus.

In Section~\ref{sec2} we give some background on previous work in torsion angle
modeling, as well as the bivariate von Mises distribution and the
Dirichlet process. In Section \ref{sec3} we present our model along with the
informative and noninformative priors. An explanation of how to fit
this model and use it for density estimation is provided in Section \ref{com:Escobar}.
Section \ref{sec5} contains an application of our method to estimate the joint
density of torsion angles in the EF loop region in the globin protein
family. Finally, we discuss our conclusions in Section \ref{sec6}.

\section{Preliminaries}\label{sec2}

We illustrate the development of our model by first exploring methods
for modeling individual torsion angle pairs.
Working with torsion angles requires the use of distributions
specifically designed to account for the behavior of angular data. This
data has the property that an angle $\phi$ is identical to the angle
$\phi+2k\pi$ for all $k\in\{\ldots,-1,0,1,\ldots\}$. The bivariate von Mises
distribution is commonly used for paired angular data.

Originally proposed as an eight parameter distribution by \citet
{Mardia1975}, subclasses of the bivariate von Mises with fewer
parameters are considered easier to work with and are often more
interpretable. \citet{Rivest1982} proposed a six parameter version,
which has been further refined into five parameter distributions. One
such subclass, known as the cosine model, was proposed by \citet
{Mardia2007}, who employed it in frequentist mixture modeling of $(\phi
, \psi)$ angles at individual sequence positions. In this paper we
consider an alternative developed by \citet{Singh2002} known as the
sine model.

The sine model density for bivariate angular observations $(\phi,\psi)$
is defined as
%
\begin{eqnarray}
\label{sine}
&&\qquad f(\phi,\psi|\mu,\nu,\kappa_1,\kappa_2,\lambda)\nonumber
\\[-8pt]\\[-8pt]
&&\qquad\qquad =C\exp\{\kappa
_1\cos(\phi
-\mu)+\kappa_2\cos(\psi-\nu)+\lambda\sin(\phi-\mu)\sin(\psi
-\nu)\}\nonumber
\end{eqnarray}
for $\phi,\psi,\mu,\nu\in(-\pi,\pi]$, $\kappa_1,\kappa_2>0$,
$\lambda\in
(-\infty,\infty)$, and
%
\begin{equation}
\label{constant}
C^{-1}=4\pi^2\sum_{m=0}^\infty\pmatrix{2m\cr m} \biggl( \frac{\lambda
^2}{4\kappa_1\kappa_2}\biggr)^m I_m(\kappa_1)I_m(\kappa_2).
\end{equation}
The parameters $\mu$ and $\nu$ determine the mean of the distribution,
while $\kappa_1$ and $\kappa_2$ are precision parameters. The parameter
$\lambda$ determines the nature and strength of association between
$\phi$ and $\psi$. This density is unimodal when $\lambda^2<\kappa
_1\kappa_2$ and bimodal otherwise. One of the most attractive features
of this particular parameterization of the bivariate von Mises is that,
when the precision parameters are large and the density is unimodal, it
can be well approximated by a bivariate normal distribution with mean
$(\mu,\nu)$ and precision matrix $\Omega$, where $\Omega
_{11}=\kappa
_1$, $\Omega_{22}=\kappa_2$ and $\Omega_{12}=\Omega_{21}=-\lambda$.

\citet{Singh2002} fit individual sine model distributions to torsion
angle data sets. \citet{Mardia2008} developed an extension of the
bivariate sine model for $n$-dimensional angular data, but the constant
of integration is unknown for $n>2$, rendering it difficult to use. We
instead consider a method based on a Dirichlet process mixture model.

The Dirichlet process, first described by \citet{Ferguson1973} and
\citet{Antoniak1974}, is a distribution of random measures which are
discrete with probability one. The Dirichlet process is typically
parameterized as having a mass parameter $\alpha_0$ and a centering
distribution $G_0$. Using the stick-breaking representation of \citet
{Sethuraman1994}, a random measure $G$ drawn from a Dirichlet process
$\mathit{DP}(\alpha_0G_0)$ takes the form
\begin{eqnarray*}
G(\mathbf{B})=\sum_{j=1}^\infty p_j\delta_{\tau_j}(\mathbf{B}),
\end{eqnarray*}
where $\delta_\tau$ is an indicator function equal to 1 if $\tau\in
\mathbf
{B}$ and 0 otherwise, $\tau_j\sim G_0$, $p_j'\sim \operatorname{Beta}(1,\alpha_0)$,
and $p_j=p_j'\prod_{k=1}^{j-1}(1-p_k')$. In this form, the discreteness
of $G$ is clearly evident.

This discreteness renders the {DP} somewhat unattractive for directly
modeling continuous data. However, it can be effectively used in
hierarchical models for density estimation [\citet{Escobar1995}].
Consider a data set $z_1,\ldots,z_n$, and a family of distributions
$f(z|\tau)$ with parameter $\tau$. A Dirichlet process mixture (DPM)
model takes the form
%
\begin{eqnarray}
\label{genericDPM}
z_i | \tau_i & \sim & f(z_i|\tau_i),\nonumber
\\
\tau_i | G & \sim & G,\nonumber
\\
G & \sim & \mathit{DP}(\alpha_0G_0).
\end{eqnarray}
The discreteness of draws from a {DP} means that there is positive
probability that $\tau_i=\tau_j$ for some $i\ne j$. For such $i$ and
$j$, $z_i$ and $z_j$ come from the same component distribution, and are
viewed as being \emph{clustered} together. The clustering induced by
DPM models generates rich classes of distributions by using mixtures of
simple component distributions.

While $\tau$ is generally taken to be scalar- or vector-valued, there
is nothing
inherent in the definition of the {DP} that imposes such a restriction,
and more complex centering distributions have been explored [e.g.,
\citet{MacEachern2000}; \citet{Iorio2004}; \citet{Gelfand2005}; \citet{Griffin2006}; \citet{Dunson2007}; \citet{Rodrigues2008}].
In a model for the distribution of multiple angle pairs, we propose
using a hidden Markov model (HMM), a discrete stochastic process, as
the centering distribution $G_0$. In the following section we describe
how to use this hidden Markov model as a component of an informative
prior for protein conformation angle data.

\section{Dirichlet process mixture model for multiple alignment
positions}\label{sec3}

The necessary Bayesian procedures to use a {DP} mixture of bivariate von
Mises sine distributions for modeling torsion angle data at individual
sequence positions were developed by \citeauthor{Lennox2009} (\citeyear{Lennox2009b,Lennox2009}). In
this section we extend this model to multiple sequence positions, and
provide a noninformative prior that directly extends the single
position model. In addition, we describe a method for using an HMM as a
centering distribution in an informative prior for sequences of
contiguous positions. We also show how to perform density estimation
using our model.

Consider a protein family data set consisting of $n$ angle pair
sequences denoted $\mathbf{x}_{1},\ldots,\mathbf{x}_{n}$. Let each observation
have $m$ sequence positions, whose angle pairs are denoted
$x_{i1},\ldots,x_{im}$ for the $i$th sequence, with $x_{ij}=(\phi
_{ij},\psi
_{ij})$. For the moment assume that we have complete data, that is,
that every $x_{ij}$ contains an observed $(\phi,\psi)$ pair. Then our
base model for the $j$th position in the $i$th sequence is as follows:
%
\begin{eqnarray}
x_{ij}  |  \theta_{ij}  &\sim&  f(x_{ij}  |  \theta_{ij}),
\nonumber
\\
\bolds{\theta}_{i} | G  &\sim&  G, \nonumber\\
 G  &\sim&  \mathit{DP}(\alpha_0 H_1H_2),
\end{eqnarray}
where $\theta_{ij}$ consists of the parameters $(\mu_{ij},\nu
_{ij},\Omega_{ij})$, $\bolds{\theta}_{i}=(\theta_{i1},\ldots,\theta_{im})$
and $f(x|\theta)$ is a bivariate von Mises sine model. The distribution
$G$ is a draw from a Dirichlet process, while $H_1$ and $H_2$ are the
centering distributions that provide atoms of the mean and precision
parameters, respectively. Note that the product $H_1H_2$ takes the role
of $G_0$ from \eqref{genericDPM}.

For our purposes, $H_2$ always consists of the product of $m$ identical
Wishart distributions we call $h_2$. This centering distribution
assumes independence for the precision parameters of sequence positions
given clustering information. Similarly, we do not assume a
relationship between the precision parameters and the mean parameters
for any sequence position, again restricting ourselves to the situation
when clustering is known. The use of a Wishart prior for bivariate von
Mises precision parameters is motivated by concerns about ease of
sampling from the prior distribution and potential issues with
identifiability. A more detailed explanation is given by \citet{Lennox2009}.

We discuss two distinct choices for $H_1$, the centering distribution
for the sequence of mean parameters $(\bolds{\mu}_{i},\bolds{\nu}_{i})$. The
first assumes {a priori} independence of the mean parameters
across sequence positions, while the second is designed to share
information across adjacent sequence positions using a hidden Markov
model based on known properties of protein secondary structure.

\subsection{Noninformative prior for multiple sequence positions}\label{noninf}

A straightforward extension of the existing single position DPM model
takes $H_1$ to be the product of $m$ identical bivariate von Mises
distributions we call $h_1$. For truly noninformative priors, a diffuse
von Mises distribution may be replaced by a uniform distribution on
$(-\pi,\pi]\times(-\pi,\pi]$. Both the von Mises and uniform versions
of the model assume {a priori} independence of the centering
parameters $(\mu_{ij},\nu_{ij})$ across sequence positions $j$.
However, dependence can still appear in the posterior distribution.
While we refer to this as the noninformative model, and use it as such,
there is no reason why informative distributions could not be used as
the components of $H_1$, nor must these components be identical. The
primary distinguishing feature of this choice of model is that no
assumptions are made as to the relationship between the mean parameters
at the various sequence positions.

An advantage of this choice for $H_1$ is that sequence positions $j$
and $j+1$ need not be physically adjacent in a protein. This situation
could be of interest when modeling the joint distribution of amino acid
residues which are not neighbors with respect to the primary structure
of a protein, but which are close together when the protein is folded.

\subsection{Informative DPM--HMM model for adjacent
sequence positions}\label{dpm-hmm}

When considering adjacent positions, however, a model assuming
independence is not making use of all available information regarding
protein structure. For this situation we recommend a centering
distribution $H_1$ that consists of a hidden Markov model incorporating
secondary structure information.

We call our model a Dirichlet process mixture on a hidden Markov model
space, or DPM--HMM. Hidden Markov models define a versatile class of
mixture distributions. An overview of Bayesian methods for hidden
Markov models is given by \citet{Scott2002}. HMMs are commonly used to
determine membership of protein families for template-based structure
modeling, but in this case the state space relates to the amino acid
sequence, also known as the primary structure [see, e.g., \citet
{Karplus1997}]. We propose instead to use an HMM for which the hidden
state space consists of the secondary structure type at a particular
sequence position. While HMMs incorporating secondary structure have
been used for {de novo} structure prediction methods [\citet
{Boomsma2008}], they have not previously been employed for
template-based strategies. We can determine both the transition
probabilities between states and the distributions of $(\phi,\psi)$
angles for each secondary structure type based on data sets in the
Protein Data Bank. Such a model provides a knowledge-driven alternative
to our noninformative prior from Section~\ref{noninf} for adjacent
sequence positions.

Our model has four hidden states corresponding to four secondary
structure metatypes defined by the Definition of Secondary Structure for
Proteins [DSSP, \citet{Kabsch1983}] program: turn (T), helix (H), strand
(E) and random coil (C). These four types are condensed from eight
basic types, with all helices being characterized as (H), $\beta$-turns
and G-turns combined into the class (T), and both strands and $\beta
$-bulges defined as (E). The model for a realization $\bolds{\theta}$ from
our hidden Markov model is defined as follows:
\begin{eqnarray*}
{\theta}_j | s_j & \sim & f(\theta_j | s_j),
\\
s_j | s_{j-1} & \sim & M(s_j | s_{j-1}),
\end{eqnarray*}
where $s_j$ defines the \emph{state} of the Markov chain at position
$j$, with $s_j\in\{1,2,3,4\}$. $M(s_j|{s_{j-1}})$ is a discrete
distribution on $\{1,2,3,4\}$ that selects a new state type with
probabilities determined by the previous state type. We set our
transition probability matrix based on 1.5 million sequence position
pairs from the PDB, while the initialization probabilities correspond
to the stationary distribution for the chain. Note that $\mathbf
{s}=(s_1,\ldots,s_m)$ is an observation from a discrete time Markov
process. We then define $f(\theta_j|s_j)$ to be a probability
distribution with parameters determined by the current secondary
structure state of the chain.

\begin{figure}
\tabcolsep=5pt
\begin{tabular}{cc}

\includegraphics{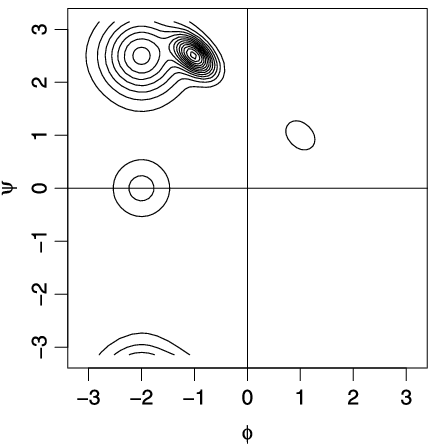}
&
\raisebox{30pt}{\begin{tabular}{r@{.}lr@{.}lr@{.}lr@{.}lr@{.}lr@{.}lr@{.}l}
\multicolumn{12}{c}{Coil Prior}\\
\multicolumn{2}{c}{$p$} & \multicolumn{2}{c}{$\mu$} & \multicolumn
{2}{c}{$\nu$} & \multicolumn{2}{c}{$\kappa_1$} & \multicolumn
{2}{c}{$\kappa_2$} & \multicolumn{2}{c}{$\lambda$}\\
0&625 & $-$2&0 & 2&5 & 4&00 & 4&00 & 0&00\\
0&208& $-$1&0 & 2&5 & 21&33 & 21&33 & $-$10&67 \\
0&125 & $-$2&0 & 0&0 & 6&25 & 6&25 & 0&00\\
0&043 & 1&0 & 1&0 & 12&21 & 12&21 & $-$3&66
\end{tabular}}
\\
\\[-8pt]

\includegraphics{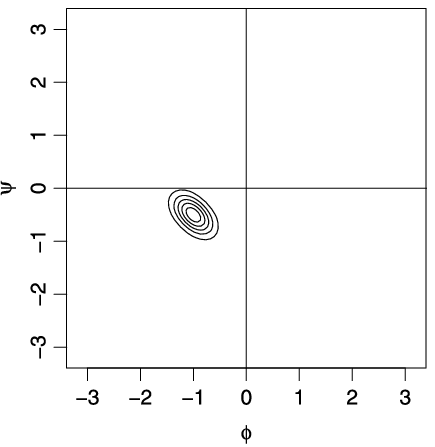}
&
\raisebox{30pt}{\begin{tabular}{r@{.}lr@{.}lr@{.}lr@{.}lr@{.}lr@{.}lr@{.}l}
\multicolumn{12}{c}{Helix Prior}\\
\multicolumn{2}{c}{$p$} & \multicolumn{2}{c}{$\mu$} & \multicolumn
{2}{c}{$\nu$} & \multicolumn{2}{c}{$\kappa_1$} & \multicolumn
{2}{c}{$\kappa_2$} & \multicolumn{2}{c}{$\lambda$}\\
 1&000 & $-$1&0 & $-$0&5 & 21&33 & 21&33 & 10&67
\end{tabular}}
\\
\\[-8pt]

\includegraphics{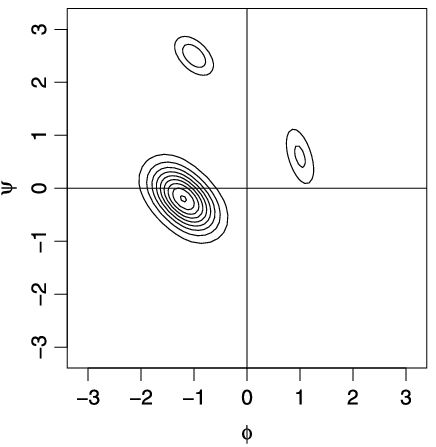}
&
\raisebox{30pt}{\begin{tabular}{r@{.}lr@{.}lr@{.}lr@{.}lr@{.}lr@{.}lr@{.}l}
\multicolumn{12}{c}{Turn Prior}\\
\multicolumn{2}{c}{$p$} & \multicolumn{2}{c}{$\mu$} & \multicolumn
{2}{c}{$\nu$} & \multicolumn{2}{c}{$\kappa_1$} & \multicolumn
{2}{c}{$\kappa_2$} & \multicolumn{2}{c}{$\lambda$}\\
0&800 & $-$1&2 & $-$0&2 & 8&33 & 8&33 & $-$4&17\\
0&100 & $-$1&0 & 2&5 & 21&33 & 21&33 & $-$10&67\\
0&100 & 1&0 & 0&6 & 33&33 & 8&33 & $-$8&33
\end{tabular}}
\\
\\[-8pt]

\includegraphics{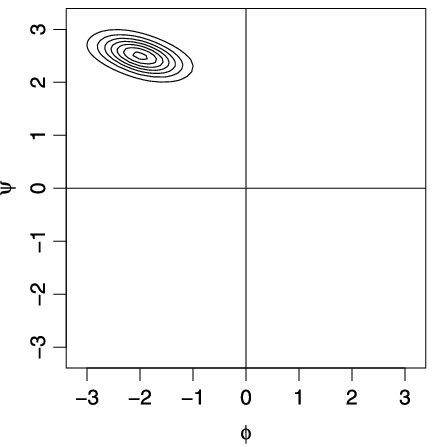}
&
\raisebox{30pt}{\begin{tabular}{r@{.}lr@{.}lr@{.}lr@{.}lr@{.}lr@{.}lr@{.}l}
\multicolumn{12}{c}{Strand Prior}\\
\multicolumn{2}{c}{$p$} & \multicolumn{2}{c}{$\mu$} & \multicolumn
{2}{c}{$\nu$} & \multicolumn{2}{c}{$\kappa_1$} & \multicolumn
{2}{c}{$\kappa_2$} & \multicolumn{2}{c}{$\lambda$}\\
1&000 & $-$2&0 & 2&5 & 5&33 & 21&33 & 5&33\\
\end{tabular}}
\end{tabular}
\caption{Graphical and numerical representations of
our von Mises mixture distributions for each of the four secondary
structure states. Note that this is the general set of secondary
structure distributions, and is not used at positions containing the
amino acids proline or glycine.}\label{fig:priors}
\end{figure}

Single bivariate von Mises distributions are not adequate to serve as
the state distributions for the four secondary structure types.
Instead, we use mixtures of between one and five bivariate von Mises
sine models. The amino acids proline and glycine exhibit dramatically
different secondary structure Ramachandran distributions, and so were
given their own distinct sets of secondary structure distributions.
Figure \ref{fig:priors} shows the state distributions used for each
secondary structure class for the eighteen standard amino acids.


%

Although these are distributions for the means of the bivariate von
Mises distribution, we chose them to mimic the distributions of $(\phi
,\psi)$ angles in each of these secondary structure classes, which
means that they are somewhat more diffuse than necessary. The use of
these secondary state distributions in conjunction with the Markov
chain on the state space allows us to leverage information about
secondary structure into improved density estimates, and provides a
biologically sound framework for sharing information across sequence positions.

Note that our model is not to be confused with the hidden Markov
Dirichlet process (HMDP) proposed by \citet{Xing2007}. The HMDP is an
implementation of a hidden Markov model with an infinite state space,
originally proposed by \citet{Beal2002}. Their model is an instance of
the Hierarchical Dirichlet Process (HDP) of \citet{Teh2006}, whereas
our DPM--HMM is a standard Dirichlet process with a novel centering distribution.

\section{Density estimation}
\label{com:Escobar}

Recall that we are interested in estimating the joint density of
$x=(\phi,\psi)$ angles at each sequence position for a candidate
structure from some protein family. Our method, as outlined by \citet
{Escobar1995}, involves treating our density estimate as a mixture of
components $f(\mathbf{x}_{n+1}|\bolds{\theta}_n+\mathrm{1})$, which in our case are
products of bivariate von Mises sine models, mixed with respect to the
posterior predictive distribution of the parameters $\bolds{\theta
}_{n+1}$. This can be written as
%
\begin{eqnarray}
f(\mathbf{x}_{n+1}|\mathbf{x}_{1},\ldots,\mathbf{x}_{n}) =\int f(\mathbf{x}_{n+1}|
\bolds
{\theta}_{n+1})\, d (\bolds{\theta}_{n+1} | \mathbf{x}_{1},\ldots,\mathbf{x}_{n}).
\end{eqnarray}
This integral cannot be written in closed form, but can be well
approximated by Monte Carlo integration. This is achieved by acquiring
samples $\bolds{\theta}^1_{n+1},\ldots,\bolds{\theta}^B_{n+1}$ from the
posterior predictive distribution for $\bolds{\theta}_{n+1}$. Then
%
\begin{equation}\label{joint}
f(\mathbf{x}_{n+1}|\mathbf{x}_{1},\ldots,\mathbf{x}_{n})\approx\frac{1}{B}\sum
_{k=1}^Bf(\mathbf{x}_{n+1}| \bolds{\theta}^k_{n+1}).
\end{equation}
While \eqref{joint} can be evaluated for any $(\phi,\psi)$ sequence
$\mathbf
{x}$, we are typically interested in graphical representations of
marginal distributions at each sequence position. For this purpose we
evaluate on a $360\times360$ grid at each alignment position. This
general Monte Carlo approach works for joint, marginal, and conditional
densities.

\subsection{Markov chain Monte Carlo}
All that remains is to determine how to obtain the samples from the
posterior predictive distribution of $\bolds{\theta}_{n+1}$, which
consists of $\bolds{\mu}_{n+1}$, $\bolds{\nu}_{n+1}$ and $\bolds{\Omega
}_{n+1}$. Fortunately, while our model is novel, the behaviors of
Dirichlet process mixtures, hidden Markov models, and the bivariate von
Mises distribution are well understood. The complexity of the posterior
distribution prevents direct sampling, but we provide the details of a
Markov chain Monte Carlo update scheme using an Auxiliary Gibbs sampler
[\citet{Neal2000}] in Appendix \hyperref[mcmc]{A}.

\subsection{The sparse data problem}
The model as described up to this point does not fully account for the
complexity of actual protein alignment data. Rather than being a simple
vector $\mathbf{x}_{i}$ of bivariate $(\phi,\psi)$ observations, the real
data also includes a vector $\mathbf{a}_{i}$ of length $m$ which consists
of variables indicating whether or not peptide $i$ was observed at each
sequence position. Let $a_{ij}=1$ if peptide $i$ is included at
alignment position $j$, and 0 otherwise. This data structure is unique
in several ways. Notice that $\mathbf{a}_{i}$ is not only known for
proteins with solved structure, but is also typically available for a
target peptide sequence. Therefore, we can avoid fitting a model that
includes alignment positions which are not of interest for our
particular problem.  This is not a true ``missing data''
problem as the unobserved sequence positions are not only absent from
our data set, but do not exist.

Our model is able to adjust to sparse data with the following
modification. Recall that the full conditional distributions could be
divided up into a prior component and a data component at each sequence
position. This makes it trivial to exclude an observation from the
likelihood, and hence posterior distribution calculation, at sequence
positions where it is not observed. For example, we can modify the full
conditional distribution of the means in the DPM--HMM model, given in
equation \eqref{hmmfc}, to be
%
\begin{eqnarray}
\qquad f(\bolds{\mu},\bolds{\nu}|\bolds{\Omega},\mathbf{x_c})\propto
L(\mathbf{s}|\bolds
{\mu},\bolds
{\nu},\mathbf{x_c})\prod_{j=1}^m f(\mu_j,\nu_j|s_j)\prod_{i\in\mathbf
{c}}f(x_{ij}|\mu_j,\nu_j,\Omega_j)^{a_{ij}}.
\end{eqnarray}
The full conditional distributions for the precision parameters and the
means with a noninformative prior, equations \eqref{scalefc} and
\eqref
{nifc}, respectively, can be modified in a similar manner. The
likelihood of $\mathbf{x}_{i}|\bolds{\theta}$, is also used by the Auxiliary
Gibbs sampler. Once again, adjust to absent data by removing unobserved
positions from the likelihood.

This model provides a straightforward method to cope with the sparse
data problem inherent in protein structure prediction. Note that the
situation in which there is ample data generally but sparse data at a
few sequence positions particularly highlights the value of the DPM--HMM
model. Secondary structure at a sparse position can be inferred based
on the surrounding positions, which can allow us to provide a better
density estimate at positions with few observed data
points.

\section{Application: Loop modeling in the globin family}\label{sec5}
\subsection{Background}

A protein's fold, or tertiary structure, consists of multiple elements
of local, regular secondary structure (repeating local motifs)
connected by the more variable loops and turns of various lengths.
These loop and turn regions can be vital to understanding the function
of the protein, as is the case in the immunoglobulin protein family
where the conformation of the highly variable loops determine how an
antibody binds to its target antigens to initiate the body's immune
response. These loop regions also tend to be the most structurally
variable regions of the protein, and modeling their structure remains
an outstanding problem in protein structure prediction [\citet
{Baker2001}]. Current knowledge-based loop modeling methods draw on
generic loop libraries. Library-based methods search the Protein Data
Bank for loops with entrance and exit geometries similar to those of
the target loop, and use these PDB loops as templates for the target
structure [e.g., \citet{Michalsky2003}]. Note that library-based methods
differ from typical template-based modeling in that they do not confine
themselves to loops within the target protein's family. Strictly within
family estimates have not previously been possible. Using the DPM--HMM
model, we are able to compare a library-based approach to a purely
within family template-based method for the EF loop in the globin family.

The globins are proteins involved in oxygen binding and transport. The
family is well studied and has many known members. Therefore, the
globin fold is suitable as a test case for template-based structure
prediction methods. A globin consists of eight helices packed around
the central oxygen binding site and connected by loops of varying
lengths. The helices are labeled A through H, with the loops labeled
according to which helices they connect. The EF loop is the longest
loop in the canonical globin structure. We generated a simultaneous
alignment of 94 members of the globin family with known tertiary
structure using MUSCLE [\citet{Edgar2004}]. For this alignment,
positions 93--106 correspond to the EF loop.

\begin{table}
\tablewidth=250pt
\caption{A table giving the details on the EF loop
for an alignment of 94 members of the globin family. The columns are
the alignment position, the number of proteins represented at the
position, the most conserved amino acid(s) at the alignment position,
and the total number of distinct amino acids observed at the alignment
position}\label{loop-basics}
\begin{tabular*}{250pt}{@{\extracolsep{4in minus 4in}}ld{2.0}cd{2.0}@{}}
\hline
\multicolumn{1}{@{}l}{\textbf{Position}} & \multicolumn{1}{c}{\textbf{\# of proteins}} & \textbf{Most conserved AA} & \multicolumn{1}{c@{}}{\textbf{\# of AAs}} \\
\hline
\phantom{0}93 & 94 & LEU & 7 \\
\phantom{0}94 & 94 & ASP & 10 \\
\phantom{0}95 & 94 & ASN & 9 \\
\phantom{0}96 & 26 & ALA & 11 \\
\phantom{0}97 & 28 & GLY & 8 \\
\phantom{0}98 & 28 & LYS & 10 \\
\phantom{0}99 & 94 & LEU & 7 \\
100 & 1 & THR & 1 \\
101 & 2 & VAL & 1 \\
102 & 2 & THR ARG & 2 \\
103 & 93 & LYS & 13 \\
104 & 94 & GLY & 15 \\
105 & 94 & ALA & 15 \\
106 & 94 & LEU & 10\\
\hline
\end{tabular*}
\end{table}

Table \ref{loop-basics} gives a summary of the behavior of 94
representative globins in the EF loop region. There is considerable
diversity in both the length and amino acid composition of this loop.
Representative loops were between 8 and 14 amino acids long, and the
highly conserved regions, particularly at the tail end of the loop,
exhibited considerable variability in amino acid composition.

We compare three different methods for loop modeling: our DPM--HMM
method with globin family data, the noninformative prior model with
globin family data, and a library-based approach. Library approaches
generate lists of loops similar to the target and use these as
templates for the target loop, generating a discrete distribution which
almost surely has mass 0 at the true conformation of the unknown loop.
To make this method comparable to our density-based approaches, we used
our noninformative prior model on library data sets to generate a
continuous density estimate. Note that all sequences in a library data
set are of the same length, which means that they will never exhibit
sparsity. For this reason, fitting the DPM--HMM model on the library
data set would not present much improvement over the noninformative model.

\subsection{Parameter settings}

For each of the 94 globins in the alignment, we generated density
estimates using each of the three methods in question. For the DPM--HMM
and noninformative models, we excluded the target from the data set
used to generate the density estimates, but used amino acid and sparse
data information from the target protein. This is reasonable since
primary structure based alignments are available for template modeling
of an unknown protein. For the library-based estimate, we applied our
noninformative prior model sequences from the coil library of \citet
{Fitzkee2005} which have the same length as the target sequence, and
have at least four sequence positions with identical amino acids.
Library data sets ranged in size from 17 to 436 angle pair sequences.

For each of our models, we ran two chains: one starting with all
observations in a single cluster and one with all observations starting
in individual clusters. Each chain was run for 11,000 iterations with
the first 1000 being discarded as burnin. Using 1 in 20 thinning, this
gave us a combined 1000 draws from the posterior distribution of the
parameters.

In all cases, our Wishart prior used $v=1$, and we set the scale matrix
$B$ to have diagonal elements of $0.25$ and off-diagonal elements of 0.
Note that we use the \citet{Bernardo1994}, pages 138--139,
parameterization, with an expected value of $vB^{-1} = B^{-1}$. Our
choice of $v$ was motivated by the fact that this is the smallest
possible value for which moments exist for the Wishart distribution,
and higher values would have lead to a more informative prior. The
choice of $B$ gave an expected standard deviation of about 30 degrees
and assumed {a priori} that there was no correlation between
$\phi
$ and $\psi$, which seemed to work well in practice. For our
noninformative prior on the means, we took $h_1$ to have $\mu_0=\nu
_0=0$, $\kappa_{10}=\kappa_{20}=0.1$ and $\lambda_0=0$. This provided
a diffuse centering distribution.

In all cases we took the {DP} mass parameter $\alpha_0$ to be 1. However,
our results were robust to departures from this value. For example, for
two randomly selected proteins we gave values for $\alpha_0$ ranging
between 0.2 and 15, giving prior expected numbers of clusters from
approximately 2--30. For our first peptide the observed mean cluster
number ranged from 3.96 to 4.46, while the second had values from 4.40
to 4.65. Thus, even our most extreme choices for the mass parameter
changed the posterior mean number of clusters by less than 1.

\subsection{Results of comparison to library}
We performed pairwise comparisons for each of our models using the
Bayes factor, defined as
%
\begin{eqnarray}
B((\bolds{\phi},\bolds{\psi}))=\frac{ f((\bolds{\phi},\bolds{\psi
})|M_1)}{f((\bolds{
\phi},\bolds{\psi})|M_2)},
\end{eqnarray}
where $M_1$ and $M_2$ are density estimates generated by two of our
three possible models. We present the results of the analyses for our
94 leave-one-out models in Table \ref{output}.

\begin{table}[b]
\tablewidth=300pt
\caption{Comparison between the DPM--HMM model on the
globin family data, noninformative prior with globin data, and
noninformative model with library data. The columns Model X and Model Y
give the percentage of the time that the likelihood for the target
conformation using Model X was greater than the likelihood of the same
conformation using Model Y. This is the equivalent to a Bayes factor
comparison with Model X in the numerator being greater than 1}\label{output}
\begin{tabular*}{300pt}{@{\extracolsep{4in minus 4in}}lcccc@{}}
\hline
\textbf{Loop length}& \textbf{Total} &\textbf{DPM--HMM} \textbf{to}& \textbf{Noninf} \textbf{to} & \textbf{DPM--HMM} \textbf{to}\\
& & \textbf{library (\%)}& \textbf{library (\%)}& \textbf{noninf (\%)}\\
\hline
\phantom{0}8 & 66 & 100 & 100 & \phantom{0}70\\
10 & \phantom{0}3 & \phantom{0}67 & \phantom{0}67 & \phantom{0}67\\
11 & 23 & 100 & \phantom{0}96 & \phantom{0}39\\
13 & \phantom{0}1 & 100 & 100 & 100\\
14 & \phantom{0}1 & 100 & 100 & 100\\
All & 94 & \phantom{0}99 & \phantom{0}98 & \phantom{0}63\\
\hline
\end{tabular*}
\end{table}

First we will address the comparison between the DPM--HMM and
noninformative models using the globin data. These models show far more
similarity to each other than to the noninformative model using the
library data, both in terms of the number of Bayes factors indicating
superiority on each side, and the fact that those Bayes factors tended
to be smaller in magnitude than those generated by comparisons to the
library models. Indeed, at positions with more than 30 observations the
marginal distributions generated by the two models appear to be very
similar. Consider the null hypothesis that the probability that the
DPM--HMM is superior to the noninformative model is less than or equal
to 0.5. A binomial test of this hypothesis yields a $p$-value of 0.009.
Of these Bayes factor results, 68 met standard criteria for substantial
evidence of superiority ($|\log_{10}(B)|>1/2$) [\citet{Kass1995}], of
which 45 supported the use of the DPM--HMM model, giving a \emph
{p}-value of 0.005. This evidence, in addition to the fact that the
combined Bayes factor, the product of all of the individual
comparisons, has a value of $10^{38}$, provides overwhelming evidence
in favor of using the DPM--HMM rather than the noninformative model. For
this reason, in the remainder of the paper, we will only refer to the
DPM--HMM model when making use of the globin data set.

Recall that the library model made use of loops of the same length as
the target, and which had a certain degree of similarity in terms of
amino acid sequence. Thus, the coil library does not exhibit any sparse
data behavior. It is also unlikely to recapture the globin family EF
loops due to the considerable variability in both length and amino acid
composition. Our results indicate that the DPM--HMM model overwhelmingly
outperforms the library-based method. Not only is the relevant Bayes
factor greater than 1 in 93 out of 94 cases, it is greater than 100 in
92 cases. The case in which the library-based method outperformed the
DPM--HMM was also significant according to the \citet{Kass1995}
criteria, so there were no ambiguous individual cases. The combined
Bayes factor was $10^{959}$, indicating that the DPM--HMM model was
definitely superior to the library overall.

\begin{sidewaysfigure}

\includegraphics{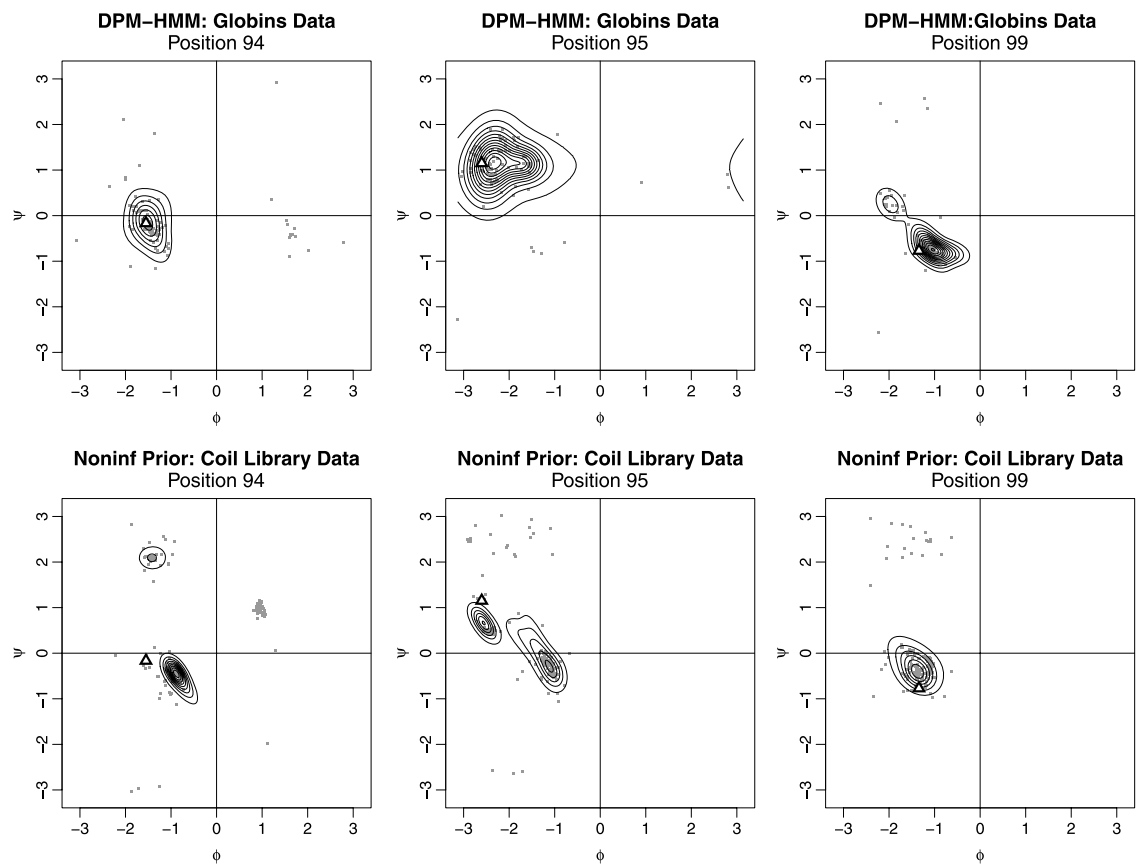}
\vspace*{-3pt}
\caption{Density estimates for positions 94, 95 and 99
for protein ``1jebD.'' The gray dots indicate the data used to fit the
model, while the triangles show the true $(\phi,\psi)$ conformation of
the target protein.}\label{loop1}
\end{sidewaysfigure}

Figure \ref{loop1} shows marginal density estimates generated for
prototypical globin ``1jebD'' for both models, along with the true
$(\phi,\psi)$ sequence for the protein for a portion of the EF loop. By
searching the PDB for loops that are similar to the target in terms of
length and sequence identity, the library method tends to place
considerable mass in areas of conformational space that are not
occupied by members of the globin family. While the members of the data
set for the globin family may not match the target loop in terms of
length or amino acid sequence, by virtue of being globins themselves
they provide a better match to the target conformation. This pattern of
improvement held true regardless of loop length. Significant
improvement was found even for the length 13 and 14 loops, for which
sparse data was a particular problem.

\subsection{Results of comparison to DBN-torus}
\label{com:DBN-torus}
\begin{figure}[b]

\includegraphics{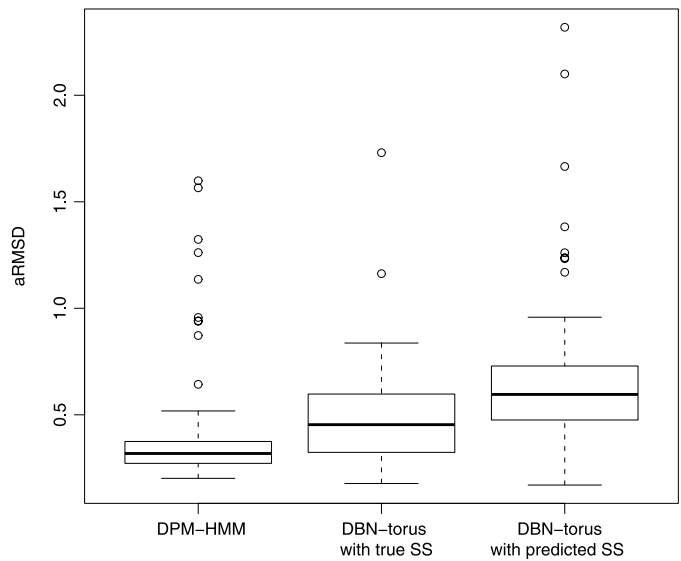}

\caption{Comparison of prediction accuracy between the
DPM--HMM and DBN-torus. DBN-torus has been given either predicted or
real secondary structure information as input. Small aRMSD values, here
given in radians, indicate predictions which are close to the target's
true tertiary structure.}\label{PHAISTOS}
\end{figure}

In addition to comparing the DPM--HMM to the knowledge-based library
method, we have also conducted a comparison to the {de novo}
DBN-torus sequence prediction method of \citet{Boomsma2008}. Unlike the
previously addressed library-based methods, DBN-torus uses continuous
density estimates, but is not customized for loop regions. It can be
used to generate sequences of predicted angle pairs given amino acid
data, secondary structure data, or no input at all. The best results
for DBN-torus are generated using amino acid data and predicted
secondary structure data. For each of our 94 targets, we generated
1000 candidate draws using the DPM--HMM, DBN-torus with predicted
secondary structure data from PsiPred [\citet{McGuffin2000}], and
DBN-torus using the true secondary structure data. Although having
exact knowledge of secondary structure for a target protein is
unrealistic in practice, it gives an idea of how well DBN-torus can
perform with optimal secondary structure prediction. We followed the
strategy of \citet{Boomsma2008} of using the angular RMSD to judge the
accuracy of our predictions. For each target, the best draw judged by
minimum aRMSD was selected, and the results are summarized in Figure
\ref{PHAISTOS}.

The DPM--HMM provides a better minimum aRMSD estimate than DBN-torus in
75$/$94 cases with predicted secondary structure information and 67$/$94
cases with true secondary structure information. Note that even under
this best case scenario, the DPM--HMM provides better predictions than
does DBN-torus. This is unsurprising, as template-based methods
typically outperform {de novo} methods where a template is
available. Proteins for which DBN-torus outperforms our DPM--HMM method
often contain an EF loop whose conformation is not a close match to
other members of the globin family. In such cases, good conformations
are more likely to be sampled from DBN-torus, which is based on the
entire PDB, rather than the DPM--HMM mimicking the behavior of the other globins.

\section{Discussion}\label{sec6}
We have presented a novel model for protein torsion angle data that is
capable of estimating the joint distribution of around 15 angle pairs
simultaneously, and applied it to extend template-based modeling to the
notoriously difficult loop and turn regions. In contrast to existing
methods such as library-based loop prediction and DBN-torus, our model
is designed to make use of only data from highly similar proteins,
which gives us an advantage when such data is available. This is a
significant advance in terms of statistical models for this type of
data, as well as a new approach to template-based structure prediction.
In addition to providing the basic model, we proposed two possible
prior formulations with interesting properties.

Our noninformative prior model, which is the direct extension of the
single position model of \citeauthor{Lennox2009} (\citeyear{Lennox2009b,Lennox2009}), provides a
method to jointly model sequence positions which may or may not be
adjacent in terms of a protein's primary structure. This model allows
for the estimation of joint and conditional distributions for multiple
sequence positions, which permits the use of innovative methods to
generate candidate distributions for protein structure.

While the noninformative prior model represents a significant advance
over existing methods, we also present an alternative model that
incorporates prior information about protein structure. This DPM--HMM
model, which uses a hidden Markov model as the centering distribution
for a Dirichlet process, uses the unique characteristics of a protein's
secondary structure to generate superior density estimates for torsion
angles at sequential alignment positions. We use a Bayes factor
analysis to demonstrate that density estimates generated with this
model are closer to the true distribution of torsion angles in proteins
than our alternative ignoring secondary structure.

Regardless of our prior formulation, the model is capable of
accommodating the sparse data problem inherent in protein structural
data, and in the case of the DPM--HMM formulation can leverage
information at adjacent sequence positions to compensate for sparse
data. This allows, for the first time, the extension of template-based
modeling to the loop regions in proteins. We show that within family
data provides superior results to conventional library and PDB-based
loop modeling methods. As loop modeling is one of the critical problems
in protein structure prediction, this new model and its ability to
enhance knowledge-based structure prediction represents a significant
contribution to this field.

Recall that our model treats the parameters of the bivariate von Mises sine
model nonparametrically through the use of the Dirichlet process prior centered
on a parametric distribution. We explored the effect of this treatment
relative to the parametric alternative of using the centering distribution
itself as the prior for the bivariate von Mises parameters. This parametric
alternative is equivalent to limiting our model to a single mixture component.
Although not every sequence position gives a strong indication of multiple
mixture components, there is at least one such sequence position for
every loop
in our data set. (See, e.g., position 94 for the coil library
data set in
Figure \ref{loop1}.) Attempts to model this data using only a single component
distribution lead to poor results, particularly since our model enforces
unimodality for each component via the Wishart prior. While the HMM
prior does allow for a mixture of bivariate von Mises distributions,
all of
these components will converge to the same distribution as the number of
observations increases, effectively reducing us to a single component
model again. The inadequacy of such a single component model is
reflected in the strong preference of the data for multiple clusters. While
the prior expected number of clusters goes to~1 as the mass parameter
$\alpha_0$ goes to 0, we found that the posterior mean number of clusters
only decreased by 1 (typically from 4 to 3) when $\alpha_0$ decreased
from 1 to
$10^{-10}$.

\begin{sidewaysfigure}

\includegraphics{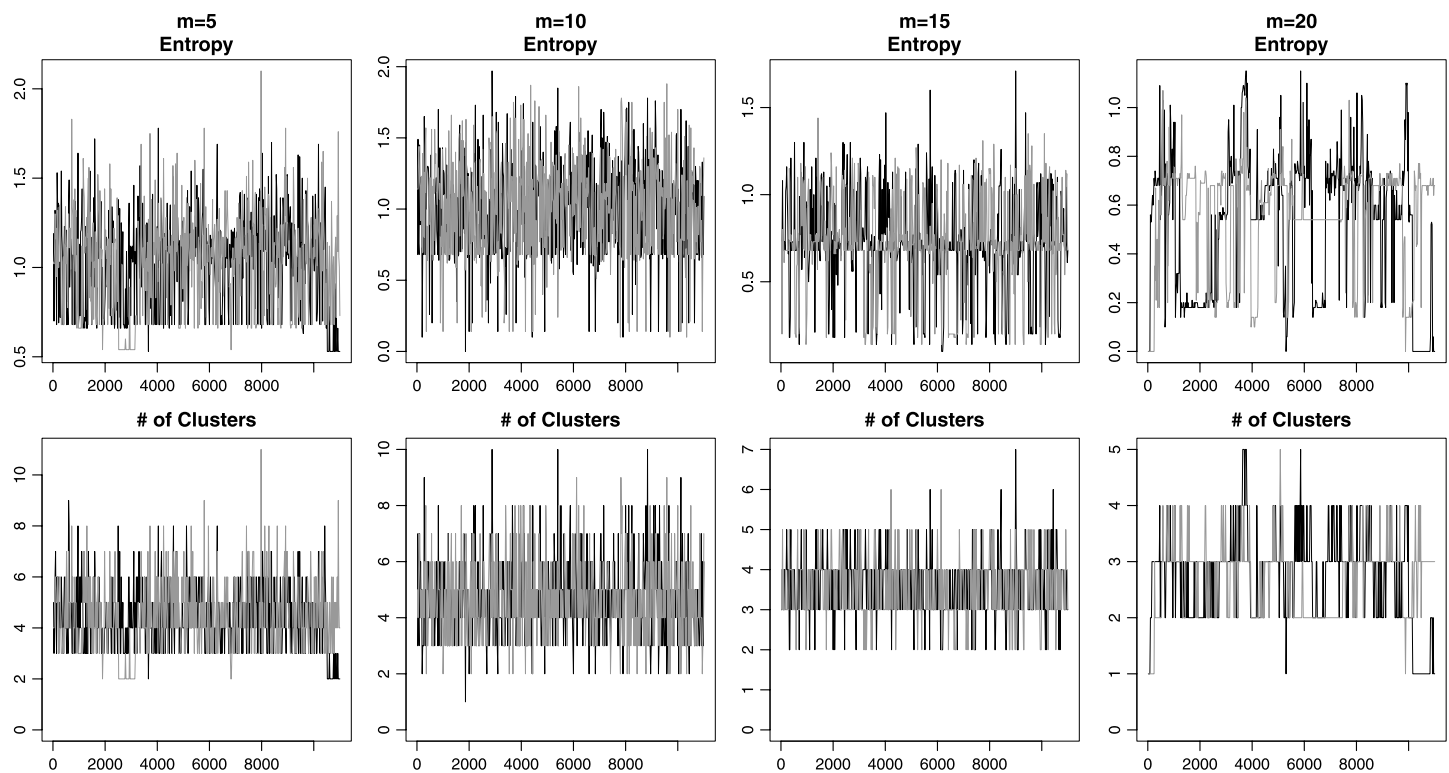}

\caption{Convergence diagnostics for density estimates
using the noninformative prior model on the globin data with contiguous
sequences beginning at position 93. Notice how mixing worsens as the
number of sequence positions increases.}\label{conv}
\end{sidewaysfigure}

In working with our sampling schemes for both the DPM--HMM and
noninformative prior models we did occasionally encounter slow mixing
and convergence problems, particularly as the number of sequence
positions under study increased. Figure \ref{conv} shows the effects on
the total number of clusters and entropy [\citet{Green2001}] per
iteration caused by increasing sequence length. As the number of
positions under study increases, there is a greater chance of getting
stuck in particular conformations, and also a subtler tendency toward
having fewer observed clusters. Although in this example the effects
are fairly mild, more severe issues can occur even at relatively short
sequence lengths. However, even when problems appear to be evident on
plots of standard convergence diagnostics, the density estimates
generated by separate chains can be quite similar. For this reason we
recommend comparing the density estimates generated by multiple chains
in addition to the standard methods of diagnosing convergence problems.

We do not recommend that our method be used for simultaneous modeling
of more than about 15 sequence positions and convergence diagnostics
should always be employed. The use of multiple MCMC chains with
different starting configurations is also highly encouraged. Particular
care should be taken with the noninformative prior model, which seems
to be more prone to these sorts of problems. We did not observe any
effect of sparse data on the speed of convergence or mixing.

Increases in sequence length and sample size both increase run time for
our software, although sequence length is the primary practical
restriction as protein families tend to have fewer than 100 members.
For the analysis of the full globins data set with 5, 10, 15 or 20
sequence positions, the run times for two chains with 11,000 iterations
using a 3 GHz processor were between 1  and 3.5 hours for the
noninformative model and 2--8 hours for the DPM--HMM.

As the emphasis in this paper is on loop modeling, which by its very
nature is limited to contiguous sequence positions, our application
does not reflect the full extent of the flexibility of our model. Our
general method is a good source of simultaneous continuous density
estimates for large numbers of torsion angle pairs. This allows us to
generate candidate models by sampling from joint distributions, or to
propagate a perturbation of the torsion angle sequence at a single
position up and down the chain through the use of conditional
distributions. Our noninformative prior model, while less impressive
than the DPM--HMM for contiguous sequence positions, can be applied to
far richer classes of torsion angle sets. This allows the modeling of
the behavior of tertiary structure motifs, which are composed of amino
acids which are not adjacent in terms of primary structure, but which
are in close contact in the natural folded state of a protein. It can
even be used to investigate the structure of polypeptide complexes, as
the $(\phi,\psi)$ positions modeled are not required to belong to the
same amino acid chain. The ability to model large numbers of $(\phi
,\psi
)$ pairs simultaneously is an exciting advance which will offer new
avenues of exploration for template-based modeling, even beyond the
field of loop prediction.

The software used in this analysis is available for download at\break
\url{http://www.stat.tamu.edu/\textasciitilde dahl/software/cortorgles/}.

\begin{table}[b]
\caption{Computational procedure}\label{minipage_procedure}
\rule{\textwidth}{0.6pt}\vspace{-18pt}
\begin{enumerate}[1.]
\item[1.] Initialize the parameter values:
\begin{enumerate}[(a)]
\item[(a)] Choose an initial clustering. Two obvious choices are: (1) one
cluster for all of the angle pair sequences, or (2) each angle pair
sequence in a
cluster by itself.
\item[(b)] For each initial cluster $\mathbf{c}$ of observed angle pair
sequences, initialize the value of the common bivariate von Mises
parameters $\bolds{\mu},\bolds{\nu},\bolds{\Omega}$ by sampling from the
centering distribution $H_1(\bolds{\mu},\bolds{\nu})H_2(\bolds{\Omega})$ of
the {DP} prior.
\begin{enumerate}[(ii)]
\item[(i)] For the noninformative prior model, sample from each of $m$
independent von Mises and Wishart distributions.
\item[(ii)] For the DPM--HMM, obtain initial values for $\bolds{\Omega}$ from $m$
independent Wishart distribution and $\bolds{\mu},\bolds{\nu}$ from the
hidden Markov model.
\end{enumerate}
\end{enumerate}
\item[2.] Obtain draws from the posterior distribution by repeating the following:
\begin{enumerate}[(a)]
\item[(a)] Given the mean and precision values, update the
clustering configuration using one scan of the Auxiliary Gibbs sampler
of \citet{Neal2000}.
\item[(b)] Given the clustering configuration and mean values, update the
precision matrix $\Omega$ for each sequence position in each cluster
using the Wishart independence sampler described in \citet{Lennox2009}.
\item[(c)] If using the DPM--HMM, obtain a draw from the full conditional
distribution of the state sequence $\mathbf{s}$ using the FB algorithm
developed by \citet{Chib1996} for each cluster.
\item[(d)] Given the clustering configuration, precision values, and (if
applicable) state information, update the values of $(\mu,\nu)$ for
each sequence position in each cluster using the independence sampler
given in Appendix \hyperref[vmmp]{B}.
\end{enumerate}
\end{enumerate}\vspace{-11pt}
\rule{\textwidth}{0.6pt}
\end{table}

\begin{appendix}
\section{Markov chain Monte Carlo}
\label{mcmc}

Here we give the details of our MCMC scheme to sample from the
posterior distribution. A concise description is provided in
Table~\ref{minipage_procedure}.
After the state of our Markov chain has been initialized, our first
step is to update the clustering associated with our Dirichlet process.
We use the Auxiliary Gibbs sampler of \citet{Neal2000} with one
auxiliary component for this purpose. Having updated the clustering, we
now must update the parameter values $\bolds{\theta}$ for each cluster by
drawing values from full conditional distribution $f(\bolds{\theta}|\mathbf
{x_c})$, where $\mathbf{x_c}=\{\mathbf{x}_i:i\in\mathbf{c}\}$ and $\mathbf{c}$ is the
set of indices for members of said cluster. Once again, this
distribution is difficult to sample from directly, so we update instead
using the full conditional distributions
$f(\bolds{\mu},\bolds{\nu}|\bolds
{\Omega
},\mathbf{x_c})$ and $f(\bolds{\Omega}|\bolds{\mu},\bolds{\nu},\mathbf{x_c})$.

In the case of the precision parameters $\bolds{\Omega}$, the full
conditional density cannot be written in closed form, but is generally
well approximated by the Wishart full conditional distribution that
results from the assumption that the data have a bivariate normal
distribution rather than a bivariate von Mises distribution. We update
$\bolds{\Omega}$ by implementing an independence sampler that uses this
``equivalent'' Wishart distribution as its proposal distribution at
each sequence position. Note that under our model, the full conditional
distribution of $\bolds{\Omega}$ does not depend on the choice of
centering distribution of the mean parameters. The full conditional is
proportional to
\begin{eqnarray}\label{scalefc}
L(\bolds{\Omega}|\bolds{\mu},\bolds{\nu},\mathbf{x_c})&\propto&
H_2(\bolds
{\Omega})
L(\mathbf{x_c}|\bolds{\Omega},\bolds{\mu},\bolds{\nu}) \nonumber
\\[-8pt]\\[-8pt]
&=& \prod_{j=1}^m h_2(\Omega_j)\prod_{i\in\mathbf{c}}f(x_{ij}|\mu
_j,\nu
_j,\Omega_j),\nonumber
\end{eqnarray}
where $h_2$ is our component Wishart prior for a single sequence
position, and $f$~is a bivariate von Mises sine model with the relevant
parameters. Notice that the positions are independent given the
clustering information, so it is trivial to update each $\Omega_j$ separately.

After updating the precision parameters at each sequence position, we
proceed to update $\bolds{\mu}$ and $\bolds{\nu}$ using an independence
sampler. For our noninformative prior, with a centering distribution
consisting of a single sine model, we use the update method described
in \citet{Lennox2009b}. In this case, with $H_1=(h_1)^n$ where $h_1$ is
a bivariate von Mises distribution, the full conditional distribution
is proportional to
\begin{eqnarray}\label{nifc}
L(\bolds{\mu},\bolds{\nu}|\bolds{\Omega},\mathbf{x_c})&\propto& H_1(\bolds{\mu
},\bolds{\nu
}) L(\mathbf{x_c}|\bolds{\Omega},\bolds{\mu},\bolds{\nu}) \nonumber
\\[-8pt]\\[-8pt]
&=& \prod_{j=1}^m h_1(\mu_j,\nu_j)\prod_{i\in\mathbf{c}}f(x_{ij}|\mu
_j,\nu
_j,\Omega_j).\nonumber
\end{eqnarray}

The DPM--HMM case where $H_1$ is defined to be a hidden Markov model is
somewhat more complicated. The positions are no longer {a priori},
and therefore {a posteriori}, independent given the clustering
information. In addition, the inclusion of an HMM in the model makes
the nature of the full conditional distribution unclear. However, if
the state chain $\mathbf{s}$ is known, draws from the full conditional are
trivial. Therefore, we rewrite our full conditional distribution, which
is proportional to
\begin{eqnarray}\label{hmmfc}
L(\bolds{\mu},\bolds{\nu}|\bolds{\Omega},\mathbf{x_c})&\propto& H_1(\bolds{\mu
},\bolds{\nu
}) L(\mathbf{x_c}|\bolds{\Omega},\bolds{\mu},\bolds{\nu}) \nonumber
\\[-8pt]\\[-8pt]
&\propto& L(\mathbf{s}|\bolds{\mu},\bolds{\nu},\mathbf{x_c})\prod_{j=1}^m f(\mu
_j,\nu
_j|s_j)\prod_{i\in\mathbf{c}}f(x_{ij}|\mu_j,\nu_j,\Omega_j),\nonumber
\end{eqnarray}
where $f(\mu,\nu|{s_j})$ is the prior distribution determined by the
state at position $j$. Recall that our priors are finite mixtures of
bivariate von Mises sine distributions. Thus, if we can generate draws
from the full conditional distribution of $\mathbf{s}$, we can update $\mu
_i$ and $\nu_i$ at each sequence position much as we did before. We use
the forward--backward (FB) algorithm of \citet{Chib1996} to sample the
full conditional distribution of $\mathbf{s}$. Note that $\mathbf{s}$ given
$\mathbf
{\mu}$ and $\bolds{\nu}$ is independent of the data. Once we have the
state information, generating samples from the distributions $\mu
_j,\nu
_j|s_j,\Omega_j,x_{\mathbf{c}j}$ is a straightforward process using an
independence sampler, the details for which are given in Appendix \hyperref[vmmp]{B}.

\section{Von Mises mixture priors}
\label{vmmp}

We present the full conditional distribution of the mean parameters
$\mu
$ and $\nu$ given that the precision matrix $\Omega$ is known and the
prior is a single bivariate von Mises distribution with parameters $\mu
_0$, $\nu_0$, $\kappa_{10}$, $\kappa_{20}$ and $\lambda_0$. Using this
information, we then prove that a finite mixture of bivariate von Mises
distributions is a conditionally conjugate prior for this model, and
present a finite mixture of sine models which serves as a good proposal
distribution.

We consider now a single sequence position, and so our data set
consists of the set $(\phi_i,\psi_i)_{i=1}^n$. The full conditional
distribution for a set of observations with bivariate von Mises sine
model distributions and a sine model prior is an eight parameter
bivariate von Mises distribution. \citet{Lennox2009b} showed that this
distribution could be represented as
\begin{eqnarray*}
f(\mu,\nu)&=&C\exp\{\tilde{\kappa}_{1}\cos(\mu-\tilde{\mu
})+\tilde{\kappa
}_{2}\cos(\nu-\tilde{\nu})
\\[2pt]
&&\qquad\ {}+[\cos(\mu-\tilde{\mu}),\sin(\mu
-\tilde{\mu
})]\tilde{A}[\cos(\nu-\tilde{\nu}),\sin(\nu-\tilde{\nu})]^T\}
\end{eqnarray*}
with parameters
\begin{eqnarray}\label{post}
\tilde{\mu}&=&\arctan\Biggl(\sum_{i=0}^n\kappa
_{1i}[\cos(\phi_i),\sin(\phi_i)]\Biggr),\nonumber
\\[2pt]
\tilde{\nu}&=&\arctan
\Biggl(\sum_{i=0}^n\kappa_{2i}[\cos(\psi_i),\sin(\psi_i)]\Biggr),\nonumber
\\[2pt]
\tilde{\kappa}_1&=&\Biggl|\sum_{i=0}^n\kappa_{1i}[\cos(\phi_i),\sin
(\phi
_i)]\Biggr|,
\\[2pt]
\tilde{\kappa}_2&=&\Biggl|\sum_{i=0}^n\kappa
_{2i}[\cos(\psi_i),\sin(\psi_i)]\Biggr|,\nonumber
\\
\qquad \tilde{A}&=& \sum_{i=0}^n\lambda_i
\left[
\matrix{
\sin(\phi_i-\tilde{\mu})\sin(\psi_i-\tilde{\nu})&-\sin(\phi
_i-\tilde{\mu
})\cos(\psi_i-\tilde{\nu})\cr
-\cos(\phi_i-\tilde{\mu})\sin(\psi_i-\tilde{\nu})&\cos(\phi
_i-\tilde{\mu
})\cos(\psi_i-\tilde{\nu})
}
\right],\nonumber
\end{eqnarray}
where $C$ is the appropriate constant of integration and the prior mean
parameters $(\mu_0,\nu_0)$ are treated as an additional observation
$(\phi_0,\psi_0)$ from a bivariate von Mises sine model with parameters
$\mu$, $\nu$, $\kappa_{10}$, $\kappa_{20}$ and $\lambda_0$.

Now consider a prior distribution of the form
\begin{eqnarray*}
\pi(\mu,\nu)&=&\sum_{k=1}^Kp_kC_k\exp\{\kappa_{10k}\cos(\mu
_{0k}-\mu
)+\kappa_{20k}\cos(\nu_{0k}-\nu)
\\
&&\qquad\hspace*{65pt} {}+\lambda_{0k}\sin(\mu
_{0k}-\mu)\sin
(\nu_{0k}-\nu)\},
\end{eqnarray*}
where $C_{k}$ is the constant of integration for a von Mises sine model
with parameters $\kappa_{10k}$, $\kappa_{20k}$ and $\lambda_{0k}$
given in equation \eqref{constant}, $p_k\ge0$ for $k=1,\ldots, K$ and
\mbox{$\sum
_{k=1}^Kp_k=1$}. The full conditional distribution is proportional to
this distribution times the likelihood, giving
\begin{eqnarray*}
&&\pi(\mu,\nu|\bolds{\phi}, \bolds{\psi})
\\
&&\qquad \propto  L(\mu,\nu|\bolds
{\phi}, \bolds
{\psi})\sum_{k=1}^Kp_kC_k
 \exp\{\kappa_{10k}\cos(\mu_{0k}-\mu
)+\kappa
_{20k}\cos(\nu_{0k}-\nu)\nonumber
\\
&&{}\hspace*{87pt}\hspace*{93pt} +\lambda_{0k}\sin(\mu_{0k}-\mu)\sin(\nu_{0k}-\nu)\} \nonumber
\\
&&\qquad =\sum_{k=1}^Kp_kL(\mu,\nu|\bolds{\phi}, \bolds{\psi})C_k
\exp\{\kappa
_{10k}\cos(\mu_{0k}-\mu)+\kappa_{20k}\cos(\nu_{0k}-\nu) \nonumber
\\
&&{}\qquad\qquad\hspace*{135pt} +\lambda_{0k}\sin(\mu_{0k}-\mu)\sin(\nu_{0k}-\nu)\},\nonumber
\end{eqnarray*}
where $L(\mu,\nu|\bolds{ \phi}, \bolds{\psi})$ is the likelihood excluding
the constant of integration.

Each term in the sum depends on the unknown parameters only through the
product of the likelihood and a single von Mises sine distribution.
This product is proportional to an eight parameter bivariate von Mises
distribution with parameters given by \eqref{post}. Call the resulting
posterior parameters $\tilde{\mu_i}$, $\tilde{\nu_i}$ and so on. Then
the full conditional distribution is proportional to
\begin{eqnarray*}
&&\sum_{k=1}^Kp_k C_k\exp\{\tilde{\kappa}_{1k}\cos(\mu-\tilde{\mu
}_k)+\tilde{\kappa}_{2k}\cos(\nu-\tilde{\nu}_k)
\\
&&\qquad \hspace*{36pt}{}+[\cos(\mu
-\tilde{\mu
}),\sin(\mu-\tilde{\mu})]\tilde{A_k}[\cos(\mu-\tilde{\mu
}),\sin(\nu
-\tilde{\nu})]^T\},
\end{eqnarray*}
which integrates to
%
\begin{eqnarray*}
\sum_{k=1}^Kp_kC_k\tilde{C}_k^{-1},
\end{eqnarray*}
where $\tilde{C}_k$ is the constant of integration for an eight
parameter bivariate von Mises distribution with parameters $\tilde{\mu
}_k$, $\tilde{\nu}_k$, $\tilde{\kappa}_{1k}$, $\tilde{\kappa}_{2k}$
and $\tilde{\lambda}_k$. Therefore, the full conditional distribution
takes the form
%
\begin{eqnarray*}
\pi(\mu,\nu|\bolds{\phi}, \bolds{\psi})=\sum_{k=1}^Kp_k^* f(\mu,\nu
|\tilde{\mu
}_k, \tilde{\nu}_k, \tilde{\kappa}_{1k}, \tilde{\kappa}_{2k},
\tilde{A}_k),
\end{eqnarray*}
where $f$ is an eight parameter bivariate von Mises distribution and
$p_k^*=(p_kC_k\tilde{C}_k^{-1})/(\sum
_{j=1}^Kp_jC_j\tilde{C}_j^{-1})$. Note that $p_k^*\ge0$ for
$k=1,\ldots,K$, and\break $\sum_{k=1}^Kp_k^*=1$.

Unfortunately computational formulas for the constant of integration of
a bivariate von Mises distribution do not exist in the general case.
Therefore, we do not sample directly from this full conditional
distribution, but rather use an independence sampler which replaces
each full conditional eight parameter distribution with a five
parameter sine model, and uses the corresponding constant of
integration from \eqref{constant}. This is accomplished by replacing
the four parameter $\tilde{A}$ with a $\tilde{\lambda}=(\sum
_{i=0}^n\lambda_ix_i^Ty_i)\{\cos(\tilde\mu-\tilde\nu
)\}
^{-1}$. [This method is a direct extension of the single sine model
prior case presented in \citet{Lennox2009b}.] Using this sampler, we
found mean and median acceptance rates around 0.52, which was
comparable to the acceptance rates for the single sine model
noninformative prior, which were around 0.55.
\end{appendix}

\section*{Acknowledgments}
The authors would like to thank J. Bradley Holmes, Jerod Parsons and
Kun Wu for help with data sets, alignments, and the torsion angle
calculations. We would also like to thank the editor, associate editor
and referees for their helpful comments.

\printaddresses

\end{document}